\documentclass[letterpaper]{article}
\usepackage{isea}
\usepackage[pdftex]{graphicx}
\usepackage{times}
\usepackage{helvet}
\usepackage{courier}
\usepackage[numbers]{natbib}

\usepackage[hyphens]{url}
\usepackage{wrapfig}
\usepackage{subcaption}
\usepackage{CJKutf8}%
\newcommand{\chinese}[1]{\begin{CJK}{UTF8}{bsmi}#1\end{CJK}} 
\newsavebox{\twosubbox} 

\usepackage{dirtytalk}
\title{Interdisciplinary Translations: Sensory Perception as a Universal Language}
\author{Xindi Kang\textsuperscript{1}, Xuanyang Huang\textsuperscript{2}, Mingdong Song\textsuperscript{3}, Varvara Guljajeva\textsuperscript{2}, JoAnn Kuchera-Morin\textsuperscript{1}
\\
\textsuperscript{1}AlloSphere Resaerch Group, Media Arts and Technology, University of California Santa Barbara, United States\\
\textsuperscript{2}Computational Media and Arts, Hong Kong University of Science and Technology, Guangzhou, China\\
\textsuperscript{3}Royal College of Art, London, United Kingdom\\
\url{xindi@ucsb.edu, xuahuang3@outlook.com, m.d.song@rca.ac.uk, varvarag@ust.hk, jkm@music.ucsb.edu}
\newline
\newline
}

\begin{document} 
\maketitle
\begin{abstract}

This paper investigates sensory perception's pivotal role as a universal communicative bridge across varied cultures and disciplines, and how it manifests its value in the study of media art, human computer interaction and artificial intelligence. By analyzing its function in non-verbal communication through interactive systems, and drawing on the interpretive model in translation studies where \say{sense} acts as a mediation between two languages, this paper illustrates how interdisciplinary communication in media art and human-computer interaction is afforded by the abstract language of human sensory perception. Specific examples from traditional art, interactive media art, HCI, communication, and translation studies demonstrate how sensory feedback translates and conveys meaning across diverse modalities of expression and how it fosters connections between humans, art, and technology. Pertaining to this topic, this paper analyzes the impact of sensory feedback systems in designing interactive experiences, and reveals the guiding role of sensory perception in the design philosophy of AI systems. Overall, the study aims to broaden the understanding of sensory perception's role in communication, highlighting its significance in the evolution of interactive experiences and its capacity to unify art, science, and the human experience.

\end{abstract}


Media Art, Interdisciplinary Communication, Interaction, Translation, Sensory Perception, HCI, Artificial Intelligence




\section{Introduction}

In recent decades, academia has witnessed the proliferation of interdisciplinary collaboration, and interdisciplinary communication has received continuous interests as a subject of investigation by researchers and scholars \cite{callaos2013interdisciplinary}. In media art, this interest manifests as practitioners face the ubiquitous presence of communication barriers when working in the cross-section of art, science and engineering. While the free flow of ideas from one discipline to another have enabled the creation of countless beautiful and impactful works, obstacles in interdisciplinary communication remain in many collaboration attempts between artists and scientists, resulting in mutual incomprehension and even prejudice \cite{Snow1959}. In this paper, we argue that sensory perception is integral to translating between disciplines and serves as a universal language in which practitioners in both media art and HCI can anchor their research. 

Shared knowledge structure facilitates the formulation of a language within a specific discipline, just like how local culture can facilitate the formulation and persistence of a specific language or dialect\cite{tanaka_persistence_2018}. Similarly, the more rigor required for a certain discipline, the more difficult it is for that discipline to integrate or translate into concepts from other areas of study. This difficulty is intrinsic in the effort to convey scientific or engineering methods and findings to the general public \cite{callaos2013interdisciplinary}. In the end, words pale in comparison to the concrete and tangible impact that scientific findings and technological advances bring to people's lives. In a similar way, artists utilize the language of art to communicate to the public. A piece of artwork can act as a non-verbal form of communication from the artist to the world. Just as Paul Brown describes in \emph{Explorations in Art and Technology}, artistic expression is \say{a powerful and uniquely human language that helps us engage with, comprehend and communicate our universe} \cite{Brown2002}.

In the domain of media art, technologies takes on new life as they become tools for artists' expression. This fusion is significantly enhanced by interactivity—when technology is woven into the artwork, allowing audiences to step into the art system and become active participants \cite{edmonds2010}.In this fusion, the role of sensory perception becomes paramount as it is the dialect through which interactive experiences speak. As technology invites audiences not just to view but to interact, their sensory engagement becomes the shared lexicon that conveys the core of the artwork, allowing art to be not only seen but felt. The intersection of HCI and interactive art, each cradled between the analytical rigor of science and the evocative power of art, highlights the transcendental function of sensory perception as a universally understood language articulating the nuance of visual expression, the practicality of utility, and the boundary-pushing nature of technological innovation. In this way, sensory perception is not just a bridge between disciplines but a unifying dialogue, harmonizing the sensory with the cognitive to create immersive experiences that resonate across the spectrum of human engagement with art and technology.

\begin{figure*}[ht]
  \begin{center}
    \includegraphics[width= 0.9\textwidth]{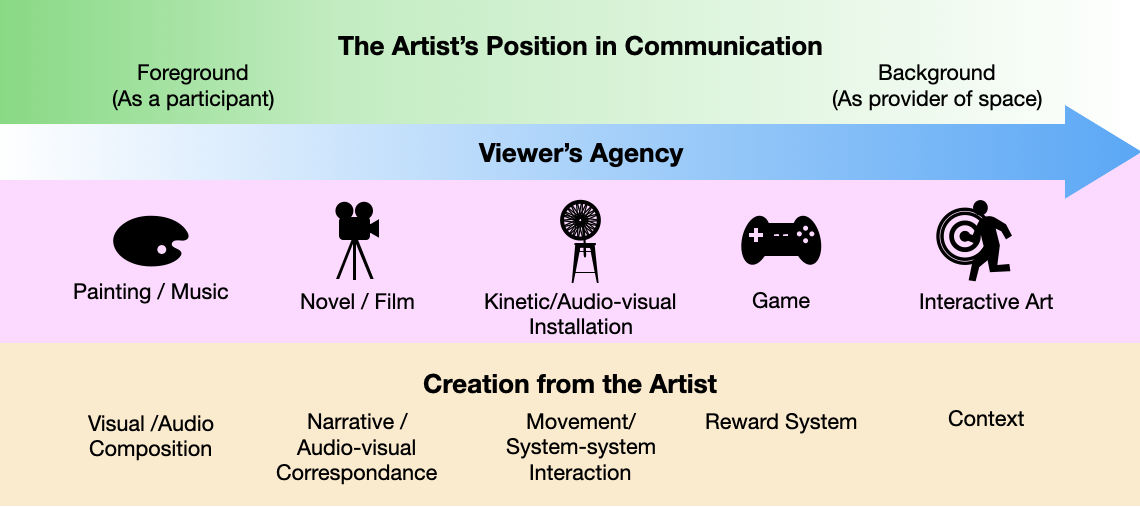}
  \end{center}
  \caption{A spectrum illustrating the artist's position in communication and the content they generate, relation to the art-form, with the degree of agency granted to viewer increasing from left to right.}
  \label{fig:spectrum}
\end{figure*}

\section{Interaction in an Interdisciplinary Context}

As media art resides at the intersection of art, science and engineering, seamless communication between disciplines is crucial for creating meaningful work in all three areas. To illustrate the integral role of sensory perception in facilitating interdisciplinary dialogue, it is crucial to begin by thoroughly examining the concept of interaction as a gateway to understanding. As the word \say{interaction} becomes the subject of study in an increasingly wide variety of fields, the meaning of the term can be quite nebulous. In the art world, for example, interaction is often considered a medium through which the artist produces an area of activity for the receivers, whose interactive actions bring to life an artwork-event \cite{Kluszczynski2010}. In interactive art, the artist is concerned with how the artwork behaves, how the audience interacts with it and, ultimately, in participant experience and their degree of engagement \cite{edmonds2010}. Whereas in engineering, the term is the central focus of research in HCI, which concerns the interaction between people (users) and computers, and the design, evaluation and implementation of user interfaces for computer systems that are receptive to the user's needs and habits \cite{Brey2009}. In psychology, interaction can then refer to the study of interpersonal activities. Psychologists are especially interested in the fundamental human nature of exploratory behavior which promotes interaction. Research in psychology suggest that participants in an interactive experience can \say{shape their experience of a realm by how they choose to move through it} and \say{alter the appearance of that realm and contribute to it}, hence the interactive experience becomes \say{a powerful psychological space} \cite{barak2008reflections}.  


\subsection{Interaction as Communication}

\begin{figure}[h]
  \begin{center}
    \includegraphics[width=0.5\textwidth]{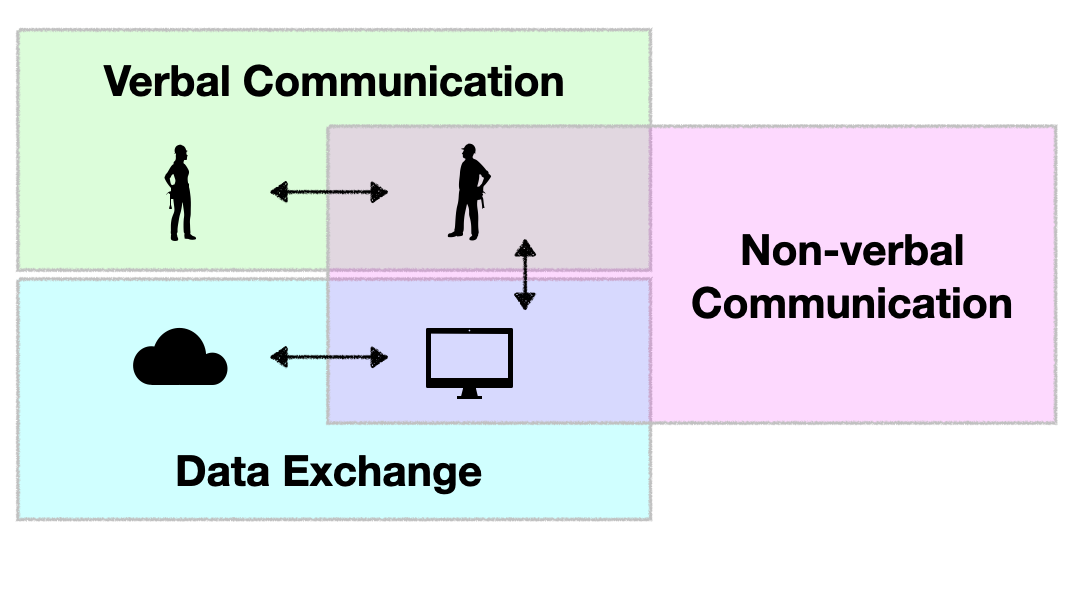}
  \end{center}
  \caption{The language used in three types of communication: Human-human, system-system, human-system}
  \label{fig:3-types-of-communication}
\end{figure}

Among the above mentioned fields of studies concerning interaction, there is an important common theme -- the study of communication. After all, interaction is most universally defined as \say{an occasion when two or more people or things communicate with or react to each other} \cite{cambridge2023}. While communication has been a long-standing subject of investigation in psychology and HCI, discussions around interactive art as a communicative medium have also been raise as early as the one Kluszczynski had initiated in his 1996 article \emph{The Context is The Message: Interactive Art as Medium for Communication}, in which the message communicated through interactive art is hypothesized to be either defined by the self-expression of the artist, or created by the recipient in the process of interaction. In his analyzis, these two modes of communication each demonstrates a unique communication mode between the artist and the viewer. In the former mode of communication, the interactive process acts as a \say{mediated interpersonal interaction with the author of the artwork} for the viewer, which he considers as closely connected to the \say{important dogmas of the modernistic aesthetic paradigm} including \say{representation, self‐expression, and the conviction about the supremacy of the artist/author’s position in the process of artistic communication}. In the latter mode, the interactive work is \say{a context in which the recipient constructs the subject matter of his/her experience and its meaning}, and the artist fades into the background as a provider of such context \cite{Kluszczynski1996}. While the claimed \say{artist's supremacy in the process of artistic communication} in the first communication mode remains debatable, given that the viewers or receivers of the message still have a certain degree of agency in their own perception of the artwork, in both cases, Kluszczynski's categorization brings up three important questions to be considered for interaction artists and designers: 
\\
\begin{itemize}
    \item Their own placements on the spectrum of communication (Fig.\ref{fig:spectrum}).
    \item The form and degree of agency the audience is granted (Fig.\ref{fig:spectrum}).
    \item The type of language used to form communication (Fig.\ref{fig:3-types-of-communication}).
\end{itemize}

In the first case brought up by Kluszczynski where the artist communicates with the viewer through the mediation of the artwork, the position of the artist is in the foreground of communication, that is, the artist is an active participant in the creation of the message, and in some cases the only one who composes the message. Typical examples of this type of communication through art are traditional art forms such as paintings, musical compositions and poems, where the viewer's agency over the form of artwork is limited. On the contrary, the second mode of communication where the artist provides a context for the viewer's creative activities, viewers gain more agency over their experience through the increased sensory channels through which interaction can occur, which forms a collaborative space in which the message is created mainly by the viewer. This is typical in interactive artworks involving active participants. In this mode, the artist's placement in the communicative activity is in the background as a provider of the space for communication to happen. This communication consists not only of interpersonal dialogue, but also an \say{intra-personal communication} which is defined as an internal dialogue between subjective I and objective Me, and a process of data transformation enabling the individual to create his/her self‐image \cite{Mead1934}. This internal dialogue, when externalized through engaging with interactive art, creates a different identity extended by the viewer's sensory connection with the virtual space within the interactive experience, which is a process Kluszczynski describes as a fragmented sense of self \cite{Kluszczynski1996}. The study of the effect of such extension or fragmentation of the sense of self through engaging in interactive art, games, and other activities in the cyberspace now forms a crucial part of the study of modern psychology \cite{barak2008reflections}.

While the above two cases illustrate an interactive system's relation to the artist's position in communication and to the agency of the audience, a separate analyzis of the types of language used to form communication in an interactive art system is needed to fully illustrate the general relationship between humans and systems. Whether it be communication between humans, systems, or within a more complex dynamic, they all follow the fundamental rule of exchange of information through language. Figure \ref{fig:3-types-of-communication} illustrates three interaction types: Human to human, system to system, and human to system interaction, each using a different type of language. 


\subsubsection{Interpersonal Interaction}

\par
Human-to-human interaction, also known as interpersonal interaction, is most effectively achieved through verbal communication within the same language and cultural system. Other than edge cases such as sign language, inter-personal communication of complex ideas and concepts can become very difficult without a mutually understandable verbal language or a translation method. However, communication between artists and viewers through an artwork can transcend cultural and language barriers through engaging in non-verbal communication such as visual compositions. Harold Cohen illustrates this point in his account of his thought process in creating AARON: 

\begin{quote}
    
    I reasoned that if an artist’s images could mean anything at all to people he had never met, it had to reflect some commonality much deeper than individual experience and probably had something to do with the fact that all human beings, regardless of their culture, use essentially the same cognitive system \cite{Cohen2002}.
    
\end{quote}

In the case of interactive artwork, especially ones involving sensory or cognitive feedback mediated through technology, this identification and utilization of the non-verbal language that pervades all human beings is an essential in the age of globalization of cultures. The practice of interactive art relies on this type of communication to evoke initial interest from the audience on the surface level, and subsequently derive deeper understanding of the artwork through sustained engagement \cite{bilda2008}.

\subsubsection{System-to-system Interaction}

\par
System-to-system interaction is a common part of the engineering and programming of computer systems as they are decomposed into a family of components that are distributed and interact by exchanging messages \cite{broy2006theory}. Simultaneously, this type of interaction is featured in many media artworks as both a principle of design and a subject for artistic creation. In the past few decades, with a plethora of interactive artworks emerging in the foreground of the media art scene featuring system-to-system interaction over human-machine interaction, it is vital to understand the language used in this form of interaction in order to gauge its influence on interactivity as a medium. 


One typical example of system-to-system interaction is demonstrated in \emph{Listening Post} by Mark Hansen and Ben Rubin. \emph{Listening Post} \cite{bullivant2005listening} is essentially an audio-visual installation which visualizes and sonifies thousands of chatroom and bulletin board conversations that are automatically collected from online in real-time. In this work, which contains virtually no interactive component for the exhibition visitors, the position of the artist can be considered to be in the foreground of communication with the audience (Fig.\ref{fig:spectrum}). However, the artists are simultaneously positioned in the background of the system-to-system communication as a provider of space for the communication to happen. While the audience at the exhibition remains mostly \say{passive}, the real-time nature of the installation also reveals an important \say{participant} in the installation - the thousands of users who engage in the chat activities displayed in the artwork. In the age of ubiquitous surveillance and frequent data breach, this form of \say{involuntary participation} is perhaps even more common in everyday life than in interactive artworks. The online activity featured in this work is typical of human-machine interaction, but it acts rather as one of the systems in the installation communicating through real-time data to another system - the large display system at the exhibition space. Erkki Huhtamo describes this setup as \say{a system that could be characterised as autonomous}. He suggests that it works \say{a bit like antique automata – even while it incorporates unpredictable elements from the Internet – the work performs certain pre-choreographed actions to the enchanted spectators kept at a distance}. In the same article he gave a few more examples of media artwork from early works such as Licht-Raum Modulator (1922-1930) by László Moholy-Nagy to Harold Cohen’s AARON and categorized them as based on the principle of \say{system interaction}, which \say{could be claimed to be the opposite of user interaction} since it \say{deliberately marginalises the active participation of the user, placing the machine and its operations in the centre} \cite{Huhtamo2004}. A few other contemporary examples include \emph{Tele-present Water} by David Bowen \cite{bowen_telepresent_water}, \emph{Dawn Chorus} by Marcus Coates \cite{coates_dawn_chorus}, and Somnium by Danny Bazo, Karl Yerkes, and Marko Peljhan \cite{somnium}, where either real-time or separately collected data from one system (natural phenomenon, human activity, and space exploration, respectively) is communicated to another system (kinetic sculpture, display groups, and robotic system, respectively).



\subsubsection{Human-to-system Interaction }

\par
Communication in human-to-system interaction utilizes a non-verbal language that can be understood by both parties, and this language is ubiquitous among HCI and interactive art. In HCI, the design of user interfaces follow a set of well-established principles to allow natural and efficient user experience, and the essence of this experience is a free-flow of information between the system and the user. As Manfred Broy indicated, an interactive system interacts with its environment by message exchange via input and output channels, and an interaction is a pattern of messages on channels \cite{broy2006theory}. Examples of such interactive systems can be traced back to the invention of the computer mouse in 1964 by Douglas Engelbart, to the recent rise of general-purpose artificial intelligence such as OpenAI's ChatGPT. Although a difference between the two methods lies in that the computer mouse interaction is a non-verbal method of communication and that of ChatGPT comes close to direct verbal interaction, in both cases, the interaction between the user and the system is based on the delivery of information through decomposition of verbal language and body action into abstract information.

In interactive art, the interaction between humans and systems follows the same rules as in HCI. Many works of art with interactive components even borrow direct forms of HCI as a medium to create interactive spaces for visitors. An example is the use of an old-style typewriter in \emph{Life Writer} by Laurent Mignonneau \& Christa Sommerer \cite{life_writer}, where typing actions result in the creation of animated artificial organisms morphed from the typed letters. In this case, communication between humans and systems is achieved through the non-verbal abstract concepts offered by the pre-defined input and output relationship of the typing interaction. However, unlike regular typing actions, the abstract concept in the piece is not only cognitively understood by the audience, but also directly felt through the senses evoked by the alterations made to the outcome of their actions. Through the animated artificial organisms, the abstract concept of content creation contained in the typing action is parsed by the sensory perception of the participant and then fed back into the interactive system, creating a sustained engagement that eventually led to deeper understanding of the language of the piece, which fits well into Edmond's model of interaction \cite{edmonds2010}.

The recent developments in interactive media, robotics and AI systems are moving human-to-system interaction increasingly closer to the uncanny valley of assimilation to interpersonal interaction, which has instilled curiosity as well as fear among the public \cite{Roose2023}. The rising interest in virtual reality systems is also telling of a tendency for interactive experiences to assimilate or even replace that of the real-world by replicating spacial awareness in virtual spaces.




\begin{figure*}[ht]
\begin{center}
\includegraphics[width=0.8\textwidth]{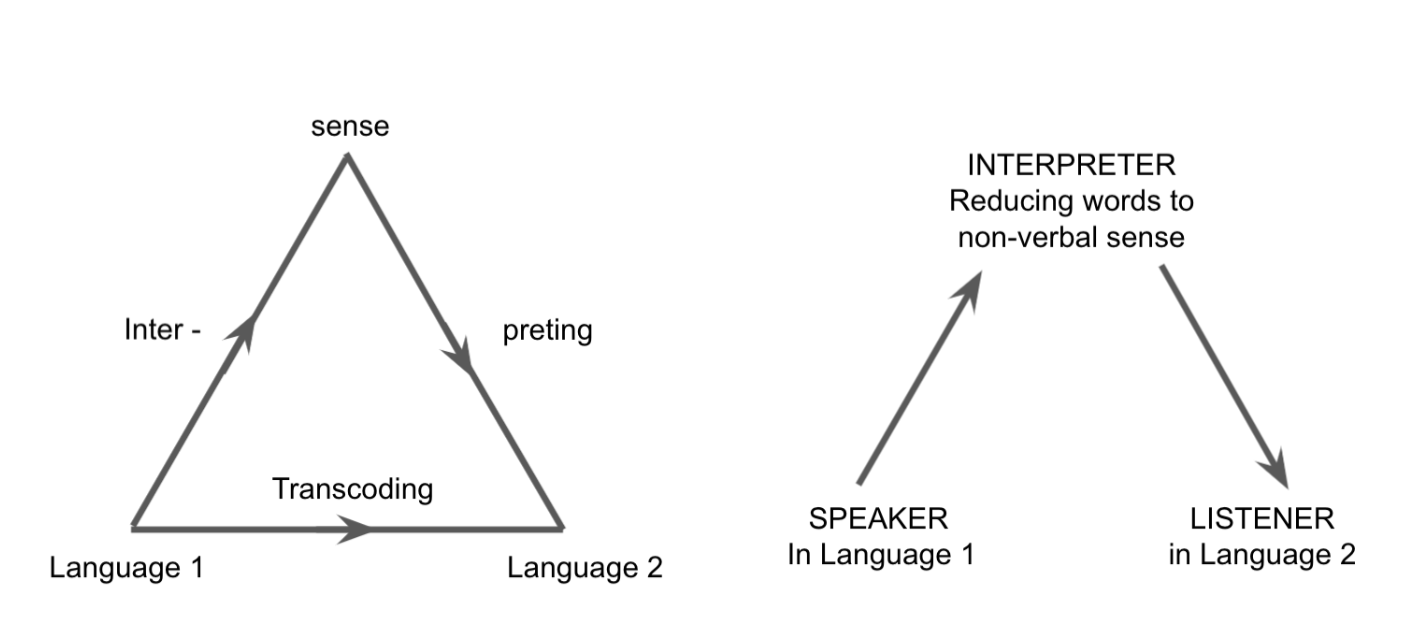}
\caption{Illustration of Seleskovitch's triangular model of ITT (two versions)}
\label{fig:seleskovitch}
\end{center}
\end{figure*}
\section{Sensory Perception: The Core of Interdisciplinary Communication}
In the Chinese version of the Heart Sutra, a scripture written around the sixth century BCE conveying the profound wisdom of Buddhist teaching, human sensory perceptions are summarized as five fundamental senses: \say{\chinese{色} (vision), \chinese{聲} (hearing), \chinese{香} (scent), \chinese{味} (taste), and \chinese{觸} (touch)}. Fifteen centuries later, the close connection between sensory perception and the human life experience remains undiminished. The embodiment of the five senses not only finds expression in artists as they create artworks, analogous patterns are discernible within the realm of technological innovation. The increasing reliance on the integration of multi-modal sensory perception systems in technological development is evident in the frequency of recent interaction and robotics design drawing inspiration from the way sensory modalities gather information and shape a person's cognition of the world through the integration of these information \cite{cao2022multimodal}. Given the multifaceted knowledge structure required for such integration, the comprehensive study and successful utilization of sensory perception in technological development is inseparable from collaboration across disciplines.

In the realm of art and communication, the concept of sensory perception takes on a unique role as a bridge between different forms of expression and understanding. This idea is exemplified by Malevich's exploration of new systems in art, wherein he advocated for the constructional inter-relationships of forms as a language of expression, transcending mere visual representation  \cite{Poltronieri2002}. This departure from representation echoes Marshall McLuhan's comparison of language to a wheel that propels intelligence forward with ease and speed  \cite{McLuhan1966}. As such, the role of sensory perception becomes paramount in facilitating communication, particularly across diverse audiences where verbal language might create barriers.

\subsection{Sensory Perception as Translation}
Sensory perception can be crucial in the process of communication, especially when information is exchanged within a vastly diverse population. When sensory perception is used as a language to establish non-verbal communication, meaning can be conveyed through abstract concepts instead of literal means. This abstraction of information automatically eliminates barriers created by differences in the language spoken, cultural background, and even age disparities among receivers. This abstraction is manifested as sensory perception is harnessed as a non-verbal language in artistic expression, where abstract concepts transcend linguistic and cultural differences. Remarkably, this notion aligns with the Interpretive Theory of Translation (ITT) put forth by Seleskovitch in 1975, which envisions translation as a process not merely between languages but also between senses and cognitive constructs \cite{seleskovitch1975}. As a pioneering theory in translation studies, ITT illuminates the importance of the underlying 'sense' that forms the basis for natural re-expression in another medium.
 

In ITT, the process of translation is a triangular process of \say{from one language to sense and from sense to the other language} \cite{pochhacker2022introducing} (figure \ref{fig:seleskovitch}). It posits that the act of translating involves de-contextualizing the source language message into the translator's cognitive understanding (sense), then re-contextualizing this understanding into the target language. Without the ITT Model, a translator approaches the text word for word, focuzing on a literal translation without considering the deeper meaning or the context. For example, the French sentence: \say{Après avoir fouillé le passé, elle a trouvé la clef} might be directly and literally translated to English as: \say{After searching the past, she found the key}. This direct method could result in a sentence that is grammatically correct but doesn't necessarily convey the metaphorical or nuanced meaning that could be implied in a more poetic or abstract source text. However, using Seleskovitch's Model, the translator first understands the meaning behind the French sentence as a 'sense' that considers the context and emotional nuances of the whole text. Suppose that \say{le passé} is a metaphor for personal history and \say{la clef} for some kind of revelation, the translation then reads as:\say{After delving into her history, she discovered the key to self-understanding}. In this Seleskovitch-influenced translation, the translator has moved beyond a literal translation to capture the sense and underlying meaning, then conveyed that in a way that is meaningful and natural in the target language. Seleskovitch's model adds depth to translation as it involves interpreting the thoughts and concepts behind the language before re-expressing them, and emphasis is placed on the ‘sense’ as a basis for natural re-expression in another language. 

Importantly, this focus on ‘sense’ manifests its value in the context of interdisciplinary translation. In the field of media art, close collaboration between artists and engineers is the ideal condition for innovation. However, in most cases the language of the artist and the language of the engineer are not translated correctly due to differences in ways of thinking. To scientists and engineers working with artistic subjects, scientific means are used to analyze artwork and the art-making process, and to create tools to assist the artist. This is particularly evident in many HCI studies that quantitatively analyze the interactive processes involved in creating art. In the world of the media artist however, science and engineering provide means of artistic expression, rather than being used to achieve a practical and utilitarian end. While this split in mindset makes it difficult for artists and engineers to collaborate, it also reveals an opportunity to find outside the box solutions. We believe that this solution is the commonality of perceptual experience. As ITT suggests that "sense" can serve as a medium of transition between two different languages, the translation between art and technology can also be realized through this medium, where one can consider the language of art and of engineering to be on the two ends of the triangle, with sensory perception acting as a common language mediating the two ends. For example, a tactile feedback experience realized through haptic engineering may have been created for the utilitarian purpose of generating stronger sensory feedback for interactive systems, but for artists this sensory feedback can still be felt in the same way when utilized in art creation. This focus on 'sense' draws parallels with cognitive processes and psychological interactions, underlining the interconnections between sensory perception and cognition, where this 'sense' can be seen as a \say{non-verbal} state \cite{pochhacker2005operation}. This non-verbal state possesses a potent capacity to impart abstract notions that defy verbal articulation, instead transmitting them through the nuances of sensory perception.


\subsection{Translation in Art Practice}

\begin{figure}[t]
  \begin{center}
    \includegraphics[width=0.45\textwidth]{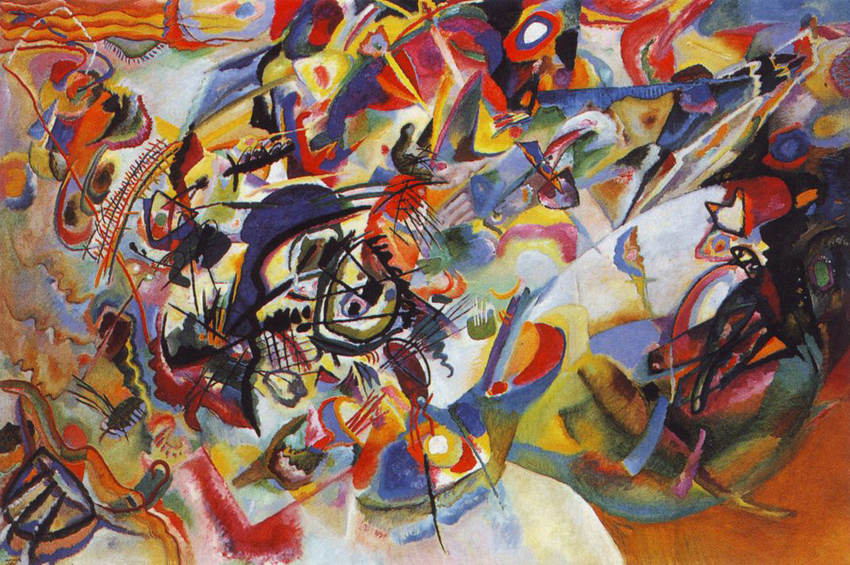}
  \end{center}
  \caption{Wassily Kandinsky,  Composition VII, 1913. Oil on canvas. \copyright Tretyakov Gallery}
  \label{fig:kandinsky}
\end{figure}

As Seleskovich's interpretive model shed light on the translation effect of sensory perception, how an artist convey meaning to the viewers through their art serves as a good example of this effect, especially when a concept is introduced through non-verbal communication engaging multiple sensory modalities. As early as the 1900s, Expressionist painter Wassily Kandinsky had already expressed in his paintings the interchangeable relationship between musical compositions and visual forms which is now studied in neuroscience as Synesthesia. For Kandinsky, music and color were inextricably tied to one another. So clear was this relationship, that Kandinsky associated each note with an exact hue. Not surprisingly, Kandinsky gave many of his paintings musical titles, such as Composition or Improvisation \cite{Miller2014} (Figure \ref{fig:kandinsky}).

\begin{figure}[t]
  \begin{center}
    \includegraphics[width=0.4\textwidth]{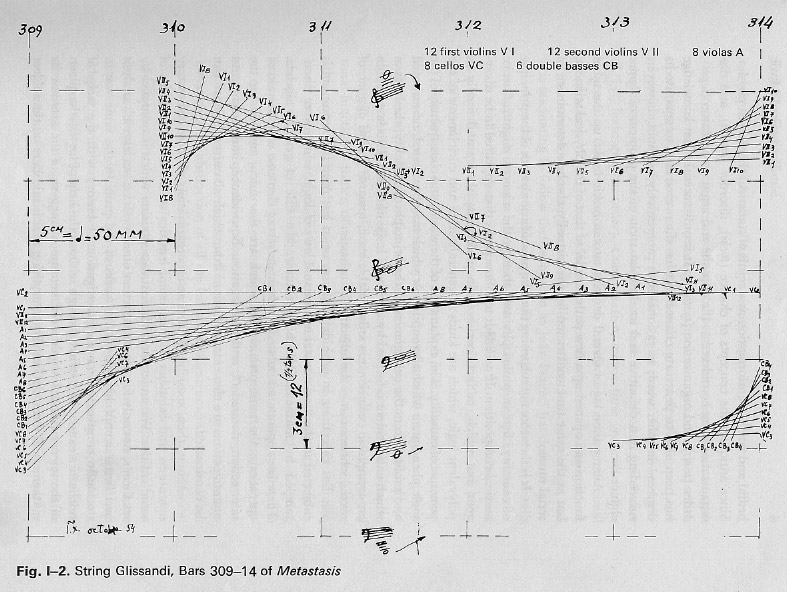}
  \end{center}
  \caption{Iannis Xenakis, Sketch showing string glissandi, mm. 309–14 \copyright Wikimedia Commons}
  \label{fig:xenakis}
\end{figure}

A contemporary of Kandinsky, Iannis Xenakis, similarly employed this cross-modal translation to convey meaning in his electronic composition \emph{Legende d'Eer}. Xenakis utilized visual scores that intertwined sonic and visual elements to elucidate the essence of his musical piece. These visual scores became a conduit for understanding the intricate auditory experience he intended to convey (Figure \ref{fig:xenakis}).

Interestingly, the significance of sensory perception extends beyond interactive art to encompass works involving system-to-system interaction. For example, David Bowen's installations, though not involving any directly interactive components, visualizations of real-time data sourced from nature, like ocean waves and flying insects, emphasize the often-overlooked sensory feedback inherent in these phenomena. 
His works visually and temporally highlight the \say{invisible} or rather \say{ignored} sensory feedback from the natural phenomena, by which the viewers are prompted to complete the experience of actually being present in such places, hence completing the feedback system that is required for interactive engagement. Similarly, Ned Kahn's large-scale kinetic sculptures, which capture and depict wave propagation patterns, integrate naturally occurring environmental patterns without relying on computational power once installed. In \emph{Wind Veil} \cite{wind_veil}, system-to-system interaction manifests as the natural real-time system of wind movement in direct contact with the kinetic sculpture system, creating for the viewer an added visual channel resulting in a heightened sensory perception of the natural phenomenon.

\subsection{Sensory Feedback Systems in Art and HCI}

As the previous section pin points sensory perception as the communicative bridge between disciplines, this section highlights the way this bridge manifests its value through media art and HCI. With sensory feedback systems establishing non-verbal communication between humans and systems, interaction becomes organically integrated into human-system interactions. This integration has garnered increasing significance in the past decade, particularly as interactive systems aim to achieve greater "smartness" or "intelligence". It is also underscored by notable instances of AI integration into interactive systems across various industries, in domains such as automated driving, sentiment-aware facial recognition, and general-purpose chat-bots. The impact of this transition is profound, as evidenced by the transformation of practical scenarios like auto-driving systems, where human sensory feedback loops inherent in driving are emulated through the utilization of depth sensors and computer vision algorithms. This emulation not only assists but also supplants human drivers in certain contexts.

Remarkably, a parallel endeavor unfolds in the realm of art. Artists, dating back to the early 2000s, have embarked on endeavors to simulate human sensory feedback systems, primarily driven by exploratory motivations. These artistic explorations have laid crucial theoretical foundations for the eventual practical application of such simulated feedback systems. An illustrative instance is found in the work of Harold Cohen, who conceived AARON, a robotic system aimed at simulating the creative processes of human painters. Cohen's meticulous attention to structuring the simulation based on feedback mechanisms inherent in diverse human behaviors—ranging from \say{operating a vehicle to guiding a forkful of food to one's mouth or executing freehand drawings—showcases the depth of his endeavor} \cite{Cohen2002}. Cohen's pioneering work in simulating sensory feedback mechanisms stands as a precursor to the subsequent surge of research into perception and cognition systems for artificial intelligence. His iterative, practice-based design approach echoes the methodologies embraced by practitioners in the realms of interactive art and human-computer interaction.
 
While both art and human-computer interaction share an inherent interest in interactive systems, the emphasis on immediacy of results within HCI research \cite{Candy2002} confers a standardized foundation to its discussions and analyzes. The amalgamation of conventional metrics with comprehensive qualitative examinations prevalent in HCI research presents an avenue to illuminate the discourse surrounding interactive art. This is particularly valuable given the categorically diverse and multifaceted nature of interactive art, often referred to as a "rhizomatic archipelago"  \cite{Kluszczynski2013} in terms of its classification. In this context, HCI researchers who strive to develop both hardware and software for artistic purposes are intrinsically motivated to comprehend the nuances of the creative process and employ artistic methodologies. 

Notably, the symbiotic relationship between practice and research is a shared characteristic of both domains \cite{Candy2002}. Consider the theory of Flow, proposed by Csikszentmihalyi (1990), which elucidates the existence of an optimal and pleasurable experiential state characterized by manageable challenges, immersion, control, freedom, lucidity, immediate feedback, temporal insensitivity, and shifts in one's self-identity \cite{Doherty2019}. This theory finds applicability in shaping the design of interactive art and human-computer systems alike, illustrating the confluence of their concerns. Moreover, the concept of engagement has been likened to a "feedback loop," in which interaction with a task molds the state-like aspects of self-efficacy and motivation, subsequently influencing a user's inclination to re-engage with the task  \cite{Doherty2019}. This insight further underscores the interplay of feedback mechanisms within the context of interactive systems.

\subsection{Navigating the Future: Sensory Perception and Artificial Intelligence}
As Alan Turing proposed in 1950 that a machine's ability to imitate human responses when verbally interacting with a human can prove its ability to think \cite{turing2012computing}, there has been heated discussion over the definition of machine \say{thinking}. Under the context of today's artificial intelligence, the definition still remains unclear. However, today's world has significantly more points of reference to discuss the machine's ability to think with terms such as \say{machine learning}, \say{computer vision}, \say{natural language processing}, \say{sentiment analyzis} and so on. From these terms an analogy can be made that the machine is being taught to see, hear, and feel, then compile the information to form appropriate responses. The fact that \say{neural networks} utilized in such learning processes are directly simulating how neurons in the human brain operate indicates that a comprehensive compilation of human sensory and cognitive experience is guiding the development of AI. It is clear that efforts are being made towards granting the machine more and more portals to perceive the world. Whether these efforts lead to a complete reproduction of the intellectual and emotional abilities of a human is debatable, just as Turing suggested when he questioned: 
\begin{quote}
    May not machines carry out something which ought to be described as thinking but which is very different from what a man does?\cite{turing2012computing}
\end{quote}
While this question may have been seen as only food for thought in Turing's time, today the intrinsic uncertainty in AI's thinking and response have raised significant concerns among the public. As interactive experiences with computing machinery become more assimilated with human-to-human communication, the tremendous power offered by AI systems adds a layer of exciting yet formidable veil to the future direction for interactive art as well as HCI. We believe that the key to navigating through this uncertainty again comes down to understanding sensory perception in both human and machine, as the human experience is ultimately connected to the deeper sensory and cognitive language that dwells in every interaction with the world. To achieve this understanding, artists and scientists must learn to speak each other's language, as the merging and transcendence beyond art and science is a must traveled road for future transformations of interactive technology as well as AI. 

In media art, art and technology borrow methods from each other, for which Kandinsky suggested an important guideline in \emph{Concerning the Spiritual in Art}:
\vspace{1mm}
\begin{quote}
``This borrowing of method by one art from another, can only be truly successful when the application of the borrowed methods is not superficial but fundamental. One art must learn first how another uses its methods, so that the methods may afterwards be applied to the borrower’s art from the beginning, and suitably. The artist must not forget that in him lies the power of true application of every method, but that that power must be developed \cite{kandinsky2012concerning}.''  
\end{quote}
\vspace{1mm}
The prejudice that hinders communication between artists and scientists, as C.P Snow pointed out in \emph{Two Cultures} in 1959 \cite{Snow1959}, can be as obstructive as the language barrier that creates mutual incomprehension. Sensory perception as a universal language which transcends cultures and disciplines can act as a catalyst for the unification of art and science. At the cross-section of art and science, media art is an important form of manifestation of such unification. As interdisciplinary research and practices transform the world, manifestations of this unification will eventually accelerate advancements in the field of media arts as well as in computer science, which in many decades may not even have the same names that they do now. The investigation into the human sensory feedback system acts as a string that connects artists, researchers, engineers and scientists to a ever-extending trajectory towards making a fundamental change in human life.
\vspace{5mm}

\bibliographystyle{isea}
\bibliography{isea}




\bibstyle{unsrt}
\end{document}